\documentclass[twocolumn,pre]{revtex4}
\usepackage{amssymb}
\usepackage{amsmath}
\usepackage{graphicx}

\begin{document}

\title{Self-assembly of DNA-coded nanoclusters}
\author{Nicholas A. Licata, Alexei V. Tkachenko}
\affiliation{Department of Physics and Michigan Center for Theoretical Physics,
University of Michigan, \ 450 Church Str., Ann Arbor, Michigan 48109}

\begin{abstract}
We present a theoretical discussion of a self-assembly scheme which makes it
possible to use DNA to uniquely encode the composition and structure of
micro- and nanoparticle clusters. These anisotropic DNA-decorated clusters
can be further used as building blocks for hierarchical self-assembly of
larger structures. \ We address several important aspects of possible
experimental implementation of the proposed scheme: the competition between
different types of clusters in a solution, possible jamming in an unwanted
configuration, and the degeneracy due to symmetry with respect to particle
permutations. \ 

\end{abstract}

\maketitle

Over the past decade, a number of proposals have identified potential
applications of DNA for self-assembly of micro- and nonostructures \cite%
{nucleic},\cite{falls},\cite{template},\cite{angstrom},\cite{nanotube}. \
Among these proposals, one common theme is finding a way to utilize the high
degree of selectivity present in DNA-mediated interactions. \ An exciting
and potentially promising application of these ideas is to use DNA-mediated
interactions to programmable self-assemble nanoparticle structures \cite%
{rational},\cite{natreview},\cite{designcrystals},\cite{nanocrystals}. \
Generically, these schemes utilize colloidal particles functionalized with
specially designed ssDNA (markers), whose sequence defines the particle
type. \ Selective, type-dependent interactions can then be introduced either
by making the markers complementary to each other, or by using linker-DNA
chains whose ends are complementary to particular maker sequences. \
Independent of these studies, there are numerous proposals to make
sophisticated nano-blocks which can be used for hierarchical self-assembly.
One recent advance in the self-assembly of anisotropic clusters is the work
of Manoharan et. al\cite{packing}. \ They devised a scheme to produce stable
clusters of $n$ polystyrene microspheres. \ The clusters were assembled in a
colloidal system consisting of evaporating oil droplets suspended in water,
with the microspheres attached to the droplet interface. \ The resulting
clusters, unique for each $n$, are optimal in the sense that they minimize
the second moment of the mass distribution $M_{2}=\sum_{i=1}^{n}(\mathbf{r}%
_{i}-\mathbf{r}_{cm})^{2}$. \ 

In this paper, we present a theoretical discussion of a method which
essentially merges the two approaches. We propose to utilize DNA to
self-assemble colloidal clusters, somewhat similar to those in Ref. \cite%
{packing}. \ An important new aspect of the scheme is that the clusters are
"decorated": each particle in the resulting cluster is distinguished by a
unique DNA marker sequence. As a result, the clusters have additional
degrees of freedom associated with particle permutation, and potentially may
have more selective and sophisticated inter-cluster interactions essential
for hierarchic self-assembly. \ In addition,\ the formation of such clusters
would be an important step towards programmable self-assembly of micro- and
nanostructures of an arbitrary shape, as suggested\ in Ref. \cite{licata}. \
\ 

\begin{figure}[h]
\includegraphics[width=2.7257in,height=2.0475in]{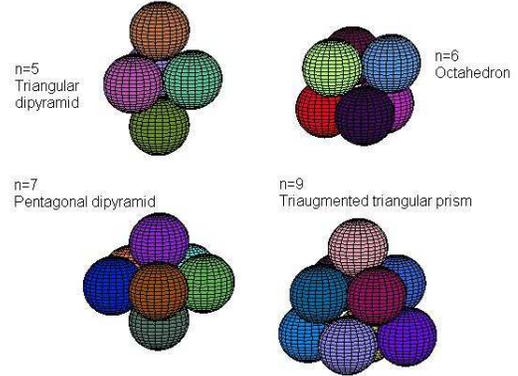}
\caption{(Color online) The minimal second moment clusters for $n=5,6,7,$
and $9$. \ Pictures of all the clusters from $n=4$ to $15$ are available in 
\protect\cite{packing}.}
\label{packingspic}
\end{figure}

We begin with octopus-like particles functionalized with dsDNA, with each
strand terminated by a short ssDNA marker sequence. \ We assume that each
particle $i$ has a unique code, i.e. the maker sequence $s_{i}$ of ssDNA
attached to it. \ We then introduce anchor DNA to the system, ssDNA with
sequence $\overline{s}_{A}\overline{s}_{B}...\overline{s}_{n}$, with $%
\overline{s}_{i}$ the sequence complementary to the marker sequence $s_{i}$.
\ The anchor is designed to hybridize with one particle of each type. \
Consider a cluster of $n$ particles attached to a single anchor. \ 

\begin{figure}[h]
\includegraphics[width=2.7257in,height=2.0517in]{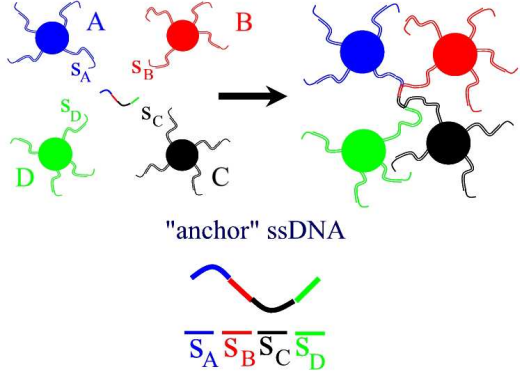}
\caption{(Color online) Schematic representation of the method for
constructing decorated colloidal clusters using ssDNA "anchors". \ }
\label{anchornew}
\end{figure}

If we treat the DNA\ which link the particles to the anchor as Gaussian
chains, there is an entropic contribution to the cluster free energy which
can be expressed in terms of the particle configuration $\{\mathbf{r}%
_{1},...,\mathbf{r}_{n}\}$ as follows. \ Here $R_{g}$ is the radius of
gyration of the octopus-like DNA arms. \ \ \ $\ $%
\begin{equation}
F=\frac{3k_{B}T}{2R_{g}^{2}}\sum_{i=1}^{n}(\mathbf{r}_{i}-\mathbf{r}%
_{anchor})^{2}
\end{equation}%
This approximation of the DNA arms as Gaussian chains is acceptable provided
their length $L$ exceeds the persistence length $l_{p}\simeq 50$ $nm$ and
the probability of self-crossing is small\cite{morphology}. \ The physical
mechanism which determines the final particle configuration in our system is
quite different from the capillary forces of Manoharan et al. \ However,
because the functional form of the free energy is equivalent to the second
moment of the mass distribution, the ground state of the cluster should
correspond to the same optimal configuration. \ 

Consider a system with $n$ particle species and an anchor of type $\overline{%
s}_{A}\overline{s}_{B}...\overline{s}_{n}$. \ The clusters we would like to
build contain $n$ distinct particles (each particle in the cluster carries a
different DNA marker sequence) attached to a single anchor. \ Let $C_{n}$
denote the molar concentration of the desired one anchor cluster. \ Because
there are many DNA attached to each particle, multiple anchor structures can
also form. \ The question is whether the experiment can be performed in a
regime where the desired one anchor structure dominates, avoiding gelation.
\ 

We consider the stability of type $C_{n}$ with respect to alternative two
anchor structures. \ To do so we determine the concentration $C_{n+1}$ of $%
n+1$ particle structures which are maximally connected, but do not have a $%
1:1:\cdots :1$ composition. \ In particular, these structures contain more
than one particle of each type, which could cause problems in our
self-assembly scheme\cite{licata}. \ There are also $n$ particle structures $%
\widetilde{C}_{n}$ with the correct composition, but which contain two
anchors. \ We would like to avoid the formation of these structures as well,
as their presence decreases the overall yield of type $C_{n}$. \ Figure \ref%
{clusternew} enumerates the various structures for an $n=3$ species system.
\ If the experiment can be performed as hoped, we will find a regime where
the ratios $\frac{C_{n+1}}{C_{n}}$ and $\frac{\widetilde{C}_{n}}{C_{n}}$\
are small. \ To this end, the equilibrium concentrations $C_{n}$, $C_{n+1}$,
and $\widetilde{C}_{n}$ are determined by equilibrating the chemical
potential of the clusters with their constituents. \ \ 

\begin{figure}[h]
\includegraphics[width=2.8853in,height=1.998in]{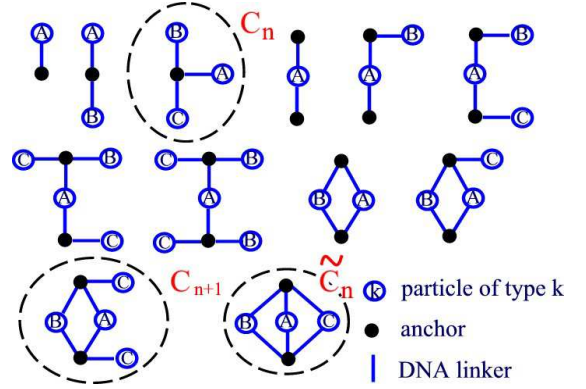}
\caption{(Color online) The topologically distinct one and two anchor
structures for an anchor ssDNA with sequence $\overline{s}_{A}\overline{s}%
_{B}\overline{s}_{C}$. \ Different structure varieites may be obtained by
relabeling the particle indices subject to the constraint that no more than
one particle of each type is attached to a given anchor. \ }
\label{clusternew}
\end{figure}

Let $c_{i}$ denote the molar concentration of species $i$, and $c_{a}$ the
molar anchor concentration. \ We consider the symmetrical case $\Delta
G_{i}=\Delta G$ and equal initial particle concentrations $%
c_{i}^{(o)}=c^{(o)}$ for all species $i$. \ In this case we have $%
c_{A}=c_{B}=...=c_{n}\equiv c$. \newline
\begin{equation}
C_{n}=c_{a}\left( Nvc\ \exp \left[ -\frac{\Delta G}{k_{B}T}\right] \right)
^{n}
\end{equation}

The binding free energy of the cluster has a contribution from the
hybridization free energy $\Delta G$ associated with attaching a particle to
the anchor, and an entropic contribution from the number of ways to
construct the cluster(since each particle has $N$ hybridizable DNA arms). \
In addition we must take into account the entropy for the internal degrees
of freedom in the structure stemming from the flexibility of the DNA
attachments to the anchor. \ In the Gaussian approximation, neglecting the
excluded volume between particles this localization volume $v=\left( \frac{%
2\pi }{3}\right) ^{\frac{3}{2}}R_{g}^{3}$ can be calculated exactly by
integrating over the various particle configurations weighted by the
Boltzmann factor: $v^{n}=\int d^{3}\mathbf{r}_{1}...d^{3}\mathbf{r}_{n}\exp %
\left[ \frac{-F}{k_{B}T}\right] $. \ 

We now consider the competing two anchor structures $C_{n+1}$ and $%
\widetilde{C}_{n}$. \ The localization volumes $v_{2}=n^{\frac{-3}{n+2}}2^{%
\frac{-3(n-1)}{2(n+2)}}v$ and $v_{3}=n^{\frac{-3}{2(n+1)}}2^{\frac{-3n}{%
2(n+1)}}v$ can be calculated in a similar fashion to $v$. \ Since there are
many DNA attached to each particle, in what follows we omit factors of $%
\frac{N-1}{N}$. \ 
\begin{equation}
C_{n+1}\simeq \frac{v_{2}^{n+2}}{v^{2n}}\frac{C_{n}^{2}}{c^{n-1}}
\end{equation}
\ 
\begin{equation}
\widetilde{C}_{n}\simeq \frac{v_{3}^{n+1}}{v^{2n}}\frac{C_{n}^{2}}{c^{n}}
\end{equation}%
The concentration of free anchors $c_{a}$ can be determined from the
equation for anchor conservation. \ \ 
\begin{equation}
c_{a}^{(o)}=c_{a}+C_{n}+2\widetilde{C}_{n}+2nC_{n+1}
\label{anchorconservation}
\end{equation}

\begin{figure}[h]
\includegraphics[width=3.1685in,height=2.3848in]{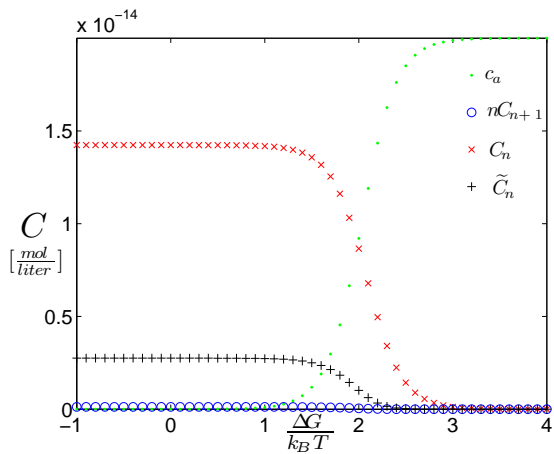}
\caption{(Color online) The molar concentrations $c_{a}$, $C_{n}$, $nC_{n+1}$%
, and $\widetilde{C}_{n}$ in the symmetrical case for a system with $n=5$
particle species. \ \ The total particle volume fraction $n\protect\phi %
\approx .25$ and $\frac{c_{a}^{(o)}}{c^{(o)}}=10^{-3}$. \ }
\label{science3new}
\end{figure}

\ \ We are interested in the low temperature regime where there are no free
anchors in solution. \ We determine the saturation values for the ratios of
interest by noting that the Boltzmann factor $\delta \equiv \exp \left[ -%
\frac{\Delta G}{k_{B}T}\right] \gg 1$ in this regime. \ 
\begin{equation}
\frac{C_{n+1}}{C_{n}}\simeq \frac{n^{-3}2^{-3n+\frac{9}{2}}}{%
(R_{g}^{3}c^{(o)})^{n-2}}\frac{c_{a}^{(o)}}{c^{(o)}}+O\left( \frac{1}{\delta
^{n}}\right)
\end{equation}

\begin{equation}
\frac{\widetilde{C}_{n}}{C_{n}}\simeq \frac{n^{-3/2}2^{-3n+\frac{3}{2}}}{%
(R_{g}^{3}c^{(o)})^{n-1}}\frac{c_{a}^{(o)}}{c^{(o)}}+O\left( \frac{1}{\delta
^{n}}\right)  \label{cntilde}
\end{equation}

Since $\frac{C_{n+1}}{\widetilde{C}_{n}}\ll 1$, eq. \ref{cntilde} provides
the experimental constraint for suppressing the two anchor structures. \
Taking the radius of the hard spheres $R\sim R_{g}$, it can be interpreted
as a criterion for choosing the initial anchor concentration $c_{a}^{(o)}$
for an $n$ species system with $\phi =\frac{4\pi }{3}R_{g}^{3}c^{(o)}$ the
particle volume fraction for an individual species. \ 
\begin{equation}
\frac{c_{a}^{(o)}}{c^{(o)}}\lesssim n^{\frac{3}{2}}(2\phi )^{n-1}
\label{criterion}
\end{equation}%
The condition gives the maximum anchor concentration for the two anchor
structures to be suppressed. \ Since $\phi \leq \frac{1}{n}$ the theoretical
limits are $\frac{c_{a}^{(o)}}{c^{(o)}}\lesssim 1$, $.29$, $.06$, and $.01$
for $n=4$, $5$, $6$, and $7$ respectively. \ In figure \ref{science3new} we
plot the solution for the concentrations. \ There is a large temperature
regime($\frac{\Delta G}{k_{B}T}\lesssim 2$) where the two anchor structures
are suppressed in favor of the desired one anchor structures. \ 

\begin{figure}[h]
\includegraphics[width=2.851in,height=2.1463in]{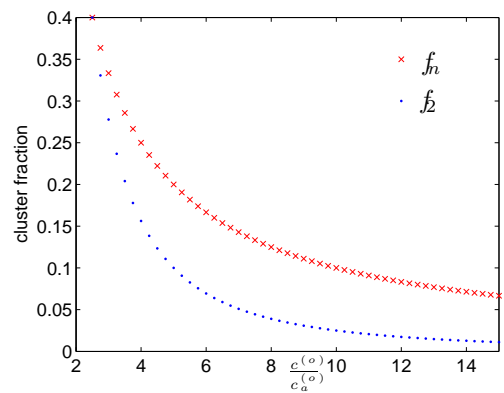}
\caption{(Color online) Cluster fractions as a function of the concentration
ratio $\frac{c^{(o)}}{c_{a}^{(o)}}$ for an $n=5$ species system in the
irreversible binding regime. \ $f_{n}$ is the fraction of the desired one
anchor structures and $f_{2}$ is the fraction of two anchor structures. \ }
\label{irreversible}
\end{figure}

The experimental criterion for suppression of the two anchor structures(eq. %
\ref{criterion}) provides a fairly strict bound on the anchor concentration,
and hence the cluster yield for $n\geq 5$. \ However, the previous
discussion considers only the equilibrium concentrations. \ Below the
melting temperature (see Figure \ref{science3new}), the connections are
nearly irreversible. \ In the irreversible regime where the DNA binding is
very strong, the probability that a cluster of type $C_{n}$ becomes attached
to a second anchor is $p\sim n\frac{c_{a}^{(o)}}{c^{(o)}}$. \ The factor of $%
n$ counts the possible attachment sites to the cluster. \ In terms of the
fraction $f_{n}\sim \frac{c_{a}^{(o)}}{c^{(o)}}$ of particles in clusters of
type $C_{n}$, the fraction of particles $f_{2}$ in the two anchor structures
will be $f_{2}\sim \frac{p}{2}f_{n}=\frac{n}{2}\left( \frac{c_{a}^{(o)}}{%
c^{(o)}}\right) ^{2}$. \ Here the factor of $\frac{1}{2}$ is necessary since
the initial attachment can be to either of the two anchors. \ As indicated
in Figure \ref{irreversible}, provided the anchor concentration is dilute
enough, $\frac{c^{(o)}}{c_{a}^{(o)}}\gtrsim 4$, the formation of the two
anchor structures is suppressed. \ 
\qquad

\begin{figure}[h]
\includegraphics[width=2.8165in,height=2.1198in]{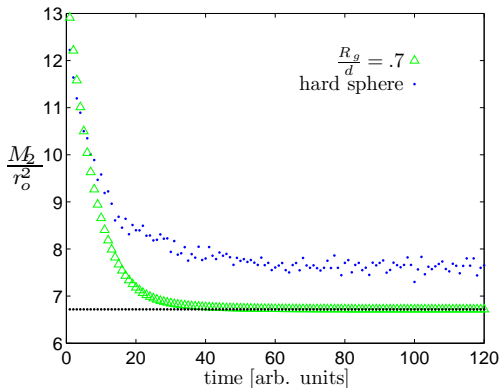}
\caption{(Color online) Plot of the dimensionless second moment $\frac{M_{2}%
}{r_{o}^{2}}$ as function of time for $n=9$ particles. \ Results are shown
for the case of hard spheres and also for a system with a soft-core
repulsion with geometric parameter $\frac{R_{g}}{d}=.7$. \ The dashed line
is the theoretical moment for the triaugmented triangular prism, which is
the minimal $n=9$ structure. \ }
\label{jampic}
\end{figure}

We now present a brief discussion of the role that jamming plays in
preventing the one anchor structures from assuming the minimal second moment
configuration. \ We performed simulations of the assembly of optimal
colloidal clusters up to $n=9$ particles by numerically integrating the
particles' Langevin equations. \ 
\begin{equation}
b^{-1}\frac{d\mathbf{r}_{i}}{dt}=-\mathbf{\nabla }_{i}H+\mathbf{\eta }_{i}
\end{equation}%
Here $b$ is the particle mobility, and the thermal noise has been
artificially suppressed(i.e. $\eta =0$). \ The model Hamiltonian $H$ used in
these simulations has been discussed in detail elsewhere\cite{licata}. \ As
indicated in Figure \ref{jampic}, the hard sphere system gets trapped in a
configuration with a larger $M_{2}$ than the optimal cluster, whereas the
soft-core system is able to fully relax. \ The jamming behavior is largely
determined by the single control parameter $\frac{R_{g}}{d}$, with $d$ the
diameter of the hard sphere. \ Beyond the critical value $\frac{R_{g}}{d}%
\gtrsim .5$ the jamming behavior is either completely eliminated, or greatly
reduced in the case of larger clusters. \ 

\begin{figure}[h]
\includegraphics[width=2.559in,height=1.9228in]{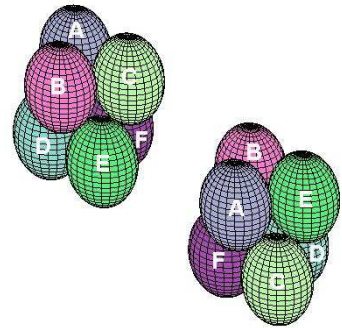}
\caption{(Color online) An illustration of degeneracy in DNA-coded
nanoclusters. \ Two different $n=6$ isomers are pictured, both with the same
minimal second moment configuration, the octahedron. \ }
\label{degeneracypicture}
\end{figure}

Building these decorated colloidal clusters is the first major experimental
step in a new self-assembly proposal. \ However, in order to utilize the
resulting clusters as building blocks, an additional ordering is necessary.
\ The problem is that the decoration introduces degeneracy in the ground
state configuration. \ This degeneracy was not present in \cite{packing}
since all the polystyrene spheres were identical. \ Namely, in the colloidal
clusters self-assembled by our method, permuting the particle labels in a
cluster does not change the second moment of the mass distribution (see
Figure \ref{degeneracypicture}). \ We need a method to select a single
"isomer" out of the many present after self-assembly. \ In the DNA-colloidal
system considered here, this isomer selection can be facilitated by "linker"
ssDNA. \ These are short ssDNA with sequence $\overline{s}_{A}\overline{s}%
_{B}$ to connect particles $A$ and $B$. \ We first construct a list of
nearest neighbors for the chosen isomer, and introduce linker DNA for each
nearest neighbor pair. \ The octopus-like DNA arms of the given particles
will hybridize to the linkers, resulting in a sping-like attraction between
the selected particle pairs. \ Note that the length $L$ of the DNA\ arms
must be on the order of the linear dimension of the original cluster. \
Otherwise the interparticle links cannot form upon introduction of linker
DNA to the system. \ It should be noted that although this method breaks the
permutation degeneracy of a cluster, the right-left degeneracy will still be
present. \ \ 

In conclusion we discussed a method which uses DNA to self-assemble
anisotropic colloidal building blocks. \ We found an experimentally
accessible regime where the resulting clusters are minimal second moment
configurations. \ In addition, the clusters are decorated: each particle in
the cluster is distinguished by a unique marker DNA sequence. \ The cluster
formation process provides an interesting model system to study a new type
of jamming-unjamming transitions in colloids. \ Constructing decorated
colloidal clusters would represent a major step towards realizing the
long-term potential of DNA-based self-assembly schemes. \ 

\textbf{Acknowledgement}. This work was supported by the ACS\ Petroleum
Research Fund (grant PRF \#44181-AC10), and by the Michigan Center for
Theoretical Physics (award \# MCTP 06-03). \ 

\bibliographystyle{achemso}
\bibliography{acompat,dna}

\end{document}